\documentstyle[12pt]{article}
\begin{document}
\title{Probing New Physics From CP Violation in Radiative B Decays }
\author{ Yue-Liang  Wu \\ Institute of Theoretical Physics, Academia Sinica, \\
 P.O. Box 2735, Beijing 100080, P.R. China } 
\date{AS-ITP-98-05}
\maketitle

\begin{abstract} 
  When new CP-violating interactions are dominated by flavor changing neutral 
  particle exchanges, that may occur in many extensions of the standard model. 
  We examine a type 3 two Higgs doublet model and find that
   direct CP asymmetries can be as large as about 25$\%$ . Time-dependent and
   time-integrated mixing-induced CP asymmetries up to 85 and 40 $\%$, 
   respectively, are possible without conflict with other constraints. 
   It mainly requirs an enhanced chromo-magnetic dipole $b\rightarrow sg$ 
   decay to be close to the present experimental bound. 
\end{abstract}

{\bf PACS numbers: 11.30.Er  12.60.Fr  13.25.Hw  13.40.Hq}

%\pacs{PACS numbers: 11.30.Er  12.60.Fr  13.25.Hw  13.40.Hq}
%\narrowtext

 The CP-averaged radiative exclusive decay $B\rightarrow K^{\ast} + \gamma$ \cite{CLEO1}
 and inclusive decay  $B\rightarrow X_{s} + \gamma$ \cite{CLEO2} represent the first 
 observation of electromagnetic penguins. The CP-averaged branching ratios 
 are consistent at the $2\sigma$ level with those \cite{TH} expected 
 from a loop involving a top quark and $W^{\pm}$ boson in the standard model (SM). 
 Within the present uncertainties from both experimental measurements and
 theoretical calculations, it is difficult, only from the CP-averaged 
 branching ratios, to distinguish the differences between the uncertainties 
 and new physics effects. However, it does not mean that contributions 
 to the amplitudes from new physics are less 
 important than the one from the SM. This is because, if there exist new 
 interactions involving new CP-violating sources, the magnitude of the 
 amplitude $A_{EMP}^{NEW}$ from new CP-violating
 interactions can become comparable with or even larger than the one $A_{EMP}^{SM}$
 from the SM even when radiative decay rate measurements are roughly that 
 expected from the SM. This happens when the contributions to the radiative decay rate 
 from the interference term $2Re(A_{EMP}^{SM}A_{EMP}^{NEW\ast})$ 
 almost cancel with the one $|A_{EMP}^{NEW}|^{2}$, i.e., $|A_{EMP}^{NEW}|^{2} \simeq 
 -2Re(A_{EMP}^{SM}A_{EMP}^{NEW\ast})$. This shows that as long as the relative
 phase between the two amplitudes ranges from about
 $(2\pi/3)$ to $(4\pi/3)$, $|A_{EMP}^{NEW}|$ will become larger than $|A_{EMP}^{SM}|$. 
 In this case, it will be alternatively interesting to investigate CP asymmetries 
 in radiative B decays to probe possible new physics effects. In ref. \cite{WW1},
 we have shown how this case occured in a type 3 two Higgs doublet model (2HDM) that was
 motivated from exploring new CP-violating sources \cite{WW2}. 
 The model can lead to a sizable direct CP asymmetry in the radiative inclusive
 decay $b\rightarrow s\gamma $ due to new CP-violating sources 
 in the charged Higgs sector. Other models (such as supersymmetric models) with possible 
 large direct CP violation were also discussed in ref.\cite{KN}. 
 In this note we will show that by simply adding
  a fourth generation of quarks and leptons in the type-3 2HDM, direct CP asymmetries in 
  radiative inclusive decays can be larger than 25$\%$. 
  The time-dependent and time-integrated mixing-induced CP asymmetries in 
  radiative exclusive decays can be as large as 100$\%$ and 50$\%$ respectively 
  due to additional new CP-violating sources via flavor changing neutral particle 
  exchanges (FCNEs). The important new contributions arise from the penguin loop 
  in which $W^{\pm}$ is replaced by the neutral scalar bosons $H^{0}_{k}$ and 
  the top quark is replaced by the fourth beauty $b'$-quark with charge $(-1/3)$. 
  One of the new interesting observations is that CP asymmetry due to the inteference 
  between penguin amplitudes via charged and neutral scalar exchanges 
  (i.e., purely new physics) can become dominant and larger than the one from 
  the interference of standard and scalar exchanges. The main reason for such large CP 
  asymmetry is due to significant chromo-magnetic dipole contributions.  
  In this case, an enhanced $b\rightarrow s g$ decay will have a branching ratio
  of order $(6-11)\%$. The values at this level appear to be favorable  
  in understanding the large branching ratio of 
  $B\rightarrow \eta' K$ decays reported recently by the CLEO Collaboration\cite{CLEO3}.

    We begin with a model independent description on CP asymmetries in radiative 
 B decays. Writting the amplitude for the processes $B\rightarrow X_{q}\gamma$ 
 ($q=s,d$) as $A_{X_{q}\gamma} = <X_{q}\gamma|O_{q\gamma}^{L}|B>
 \hat{C}_{q\gamma}^{L} + <X_{q}\gamma|O_{q\gamma}^{R}|B>\hat{C}_{q\gamma}^{R}$ with 
 $O_{q\gamma}^{L,R}  = -(G_{F}m_{b}/4\sqrt{2} \pi^{2}) e \bar{q} \sigma^{\mu\nu} 
P_{L,R}b F_{\mu\nu}$ with $P_{L,R} = (1\pm \gamma_{5})/2$, the coefficients $\hat{C}_{q\gamma}^{L,R}$ can 
be expressed by the following form
\begin{eqnarray}
\hat{C}_{q\gamma}^{L,R} & = & v_{t}^{q}C_{q\gamma}^{L,R} +
 i v_{t}^{q}C_{qg}^{L,R}f_{g\gamma}  \nonumber \\
 & + & i\left( v_{c}^{q}C_{qc}^{L,R}f_{c\gamma} 
 + v_{u}^{q}C_{qu}^{L,R}f_{u\gamma} \right) 
 \end{eqnarray}
For the charge conjugate processes $\bar{B}\rightarrow X_{\bar{q}}\gamma$, one has 
\begin{eqnarray}
\hat{C}_{\bar{q}\gamma}^{L,R}  & = & v_{t}^{q\ast}C_{q\gamma}^{L,R\ast} +
 i v_{t}^{q\ast}C_{qg}^{L,R\ast}f_{g\gamma}  \nonumber \\
 & + & i\left( v_{c}^{q\ast}C_{qc}^{L,R\ast} f_{c\gamma} 
 + v_{u}^{q\ast}C_{qu}^{L,R\ast}f_{u\gamma} \right) 
\end{eqnarray}
where $v_{i}^{q} = V^{\ast}_{iq}V_{ib} $ (i=t,c,u) with $V_{ij}$ being 
the CKM matrix elements. $C_{qg}^{L,R}$ and $C_{qf}^{L,R}$ (f=c,u) are Wilson 
coefficient functions for gluon emission and four quark operators. 
$f_{q\gamma}$, $f_{c\gamma}$ and $f_{u\gamma}$
 represent virtual corrections to the matrix element due to final state interactions 
or rescatterings\cite{JMS,GHW}. 

 To see how the CP asymmetry observable $A_{X_{q}\gamma}^{CP}$ is sensitive to 
 new physics, we may decompose $C_{qv}^{L,R}$ ($v=\gamma, g$) and $C_{qf}^{L,R}$ 
 $(f=u,c)$ into two parts: $C_{qv}^{L,R} = \bar{C}_{qv}^{L,R} + \tilde{C}_{qv}^{L,R}/v_{t}^{q}$ 
 and $C_{qf}^{L,R} = \bar{C}_{qf}^{L,R} + \tilde{C}_{qf}^{L,R}/v_{t}^{q}$.
Where $\bar{C}_{ij}^{L,R} $ denote the contributions from the SM and 
$\tilde{C}_{ij}^{L,R}$ represent the corrections from possible new physics. In the SM, 
one has $\bar{C}_{qv}^{L,R} = \sum_{f}[\bar{C}_{qv}^{L,R}(f)v_{f}^{q}/v_{t}^{q}$]($f=t,c,u$)
with $\bar{C}_{qv}^{L,R}(f)$ are Wilson coefficient functions corresponding to 
a loop-quark $f$. As the photon in the SM is predominantly left-handed and $m_{u}, m_{c} \ll
m_{t}$,  to a good approximation, we will ignore the contributions from coefficient 
functions $\bar{C}_{qv}^{R}$ and $\bar{C}_{qv}^{L}(f)$ with $f=c,u$. It is also expected
that new physics effects at the tree level are small in comparison with the one in the SM, 
namely, $|\tilde{C}_{qf}^{L}|\ll |\bar{C}_{qf}^{L}| $ and 
$|\tilde{C}_{qf}^{R}| \ll 1$ with $f=c,u$. With these considerations,  
CP asymmetry $A_{X_{q}\gamma}^{CP}$ is given by
\begin{eqnarray}
& & A_{X_{q}\gamma}^{CP} = \frac{\Gamma - \bar{\Gamma}}{ \Gamma + \bar{\Gamma}}
 \simeq  \frac{2}{ |C_{q\gamma}^{L}|^{2} +|C_{q\gamma}^{R}|^{2} }   \nonumber \\
& & \cdot [ f_{g\gamma} \left( \bar{C}_{qg}^{L}(t) Im (\tilde{C}_{q\gamma}^{L}/v_{t}^{q})
- \bar{C}_{q\gamma}^{L}(t) Im (\tilde{C}_{q g}^{L}/v_{t}^{q}) \right) \nonumber \\
 & & +  \left(\bar{C}_{qc}^{L}f_{c\gamma} Im(\tilde{C}_{q\gamma}^{L}v_{c}^{q\ast}) 
 + \bar{C}_{qu}^{L}f_{u\gamma} Im(\tilde{C}_{q\gamma}^{L}v_{u}^{q\ast})  
 \right)/|v_{t}^{q}|^{2} \nonumber \\
 & & +  f_{g\gamma} \left( Im (\tilde{C}_{q\gamma}^{L}\tilde{C}_{qg}^{L\ast})
 + Im (\tilde{C}_{q\gamma}^{R}\tilde{C}_{qg}^{R\ast}) \right)/|v_{t}^{q}|^{2} \\
& & -  \bar{C}_{q\gamma}^{L}(t) \left( \bar{C}_{qc}^{L}f_{c\gamma} Im(v_{c}^{q}/v_{t}^{q})
 + \bar{C}_{qu}^{L}f_{u\gamma} Im(v_{u}^{q}/v_{t}^{q}) \right) ]  \nonumber 
 \end{eqnarray}
where the last term is the pure standard model contributions and the third term represents pure
new physics effects. The first two terms arise from the interference between the SM and new
CP-violating interactions. Here the first term is related to the chromo-magnetic dipole operator
and the second term to the four quark operators. For simplicity, we will not consider 
contributions from gluon bremstrahlung processes as they do not affect  our main conclusions.

  To demonstrate which CP-violating sources may lead to a significant CP asymmetry in the 
  radiative B decays, we first consider the case that only one of new CP-violating interactions
  plays a dominant role. In this case the third term in the bracket of eq.(3) vanishes since 
  $\tilde{C}_{q\gamma}^{L,R}$ and $\tilde{C}_{qg}^{L,R}$ have the same phase.  For the 
  decays $B\rightarrow X_{s} + \gamma$, CP asymmetry from the SM is below $1\%$\cite{JMS}. 
  The second term in the bracket of eq.(3) is suppressed either from the CKM matrix element 
  $|v_{u}^{s}/v_{t}^{s}|\ll 1$ or from phase space factor that was found \cite{GHW} 
  at the leading-order to be $f_{c\gamma} \simeq 0.145f_{u\gamma}\simeq 0.16 f_{g\gamma}$ and 
  $f_{g\gamma} = 2\alpha_{s}/9$ \cite{AI}.  In this case,  possible 
  significant CP asymmetry in the decays $B\rightarrow X_{s} + \gamma$ will mainly arise 
  from the first term in the bracket of eq.(3). We further consider the case that 
  $|C_{q\gamma}^{R}|^{2}\ll |C_{q\gamma}^{L}|^{2}$ and 
  will focus on the first term in the bracket of eq.(3).  Its maximal value occurs at 
  $\left(\bar{C}_{q\gamma}^{L} + Re(\tilde{C}_{q\gamma}^{L}/v_{t}^{q})
  \simeq 0\right)$ with $Im(\tilde{C}_{q\gamma}^{L}/v_{t}^{q}) \simeq \pm 
  |\bar{C}_{q\gamma}^{L}|$ even when radiative decay rate measurements agree with the 
  SM predictions. One then has for this case 
\begin{equation}
a_{X_{q}\gamma}^{CP}|_{max} \simeq \mp \frac{4\alpha_{s}}{9} 
\left(\frac{\bar{C}_{qg}^{L}(t)}{\bar{C}_{q\gamma}^{L}(t)} -
 \frac{Im(\tilde{C}_{q g}^{L}/v_{t}^{q})}{Im(\tilde{C}_{q\gamma}^{L}/v_{t}^{q})} \right)
\end{equation} 
where $\bar{C}_{q\gamma}^{L}(t)$ and $\bar{C}_{qg}^{L}(t)$ have been evaluated to 
the next-to-leading order corrections\cite{TH}. In our present considerations, 
only leading-oder corrections\cite{LO} are needed.  
When evaluating with $\alpha_{s}(m_{b}) \simeq 0.214$ and $m_{t}(m_{t}) = 175$GeV, 
we have  $\bar{C}_{q\gamma}^{L}(t) \simeq -0.31$  and $\bar{C}_{qg}^{L}(t)\simeq -0.15$. 

 To apply the above model-independent analyses to extensions of the SM, we first 
 consider the simple type-3 2HDM presented in \cite{WW1,WW2}. The important new contributions to  
 penguin loops arise from new CP-violating sources in the charged Higgs sector. In that case,  
 we have $Im(\tilde{C}_{q v}^{L}/v_{t}^{q}) = Im(\chi_{q}^{H^{+}}) C_{q v}^{H^{+}} $ 
 ($v=\gamma, g$). Here $\chi_{q}^{H^{+}} = \xi_{t}\xi_{b}$ 
 with $\xi_{i}$ being complex couplings \cite{WW1}. 
$C_{q\gamma}^{H^{+}}$ and $C_{qg}^{H^{+}}$ are integral functions\cite{WW1}.
For charged Higgs mass ranging from $m_{H^{+}} \simeq 100$ GeV to $m_{H^{+}} \simeq 1$ TeV, 
we find that the corresponding values of the maximal CP asymmtry range from 
$a_{X_{q}\gamma}^{CP}|_{max} \simeq \pm 5\%$ to 
$a_{X_{q}\gamma}^{CP}|_{max} \simeq \pm 10\%$. It is of interest to note that the 
 values of the above defined maximal CP asymmetry increase as the charged Higgs mass goes up.
This is because the ratio $ C_{qg}^{H^{+}}/C_{q\gamma}^{H^{+}}$ increases as the 
charged Higgs mass becomes larger. 

 We now consider another interesting case in which new CP-violating interactions are dominated by 
 flavor changing neutral particle exchanges (FCNEs). This case may occur in many extensions of 
 the SM. As a simple example, we further consider the type-3 2HDM by adding a fourth
 generation of quarks and leptons with heavy neutrinos. It has been pointed out by Cheng and 
 Sher\cite{CS} and others and reemphasized by Hall and Weinberg \cite{HW} that FCNE may be 
 suppressed by an approximate flavor symmetry. The scalar interactions concerned in 
 our present considerations are given in the physics basis by 
 \begin{eqnarray}
 L_{S} & = & (\sqrt{2}G_{F})^{1/2} \sum_{k=1}^{3}H_{k}^{0}[
 S^{(k)}_{bb'}\bar{b}_{L}b'_{R} + 
 S^{(k)}_{b'b}\bar{b'}_{L}b_{R}  \nonumber \\
 & + &  \sum_{q=s,d}(S^{(k)}_{qb'}\bar{q}_{L}b'_{R} + S^{(k)}_{b'q}\bar{b'}_{L}q_{R}) + h.c ]
 \end{eqnarray}
 with $S^{(k)}_{ij}$ being parameterized by the averaged masses 
 $S^{(k)}_{ij}= \sqrt{m_{i}m_{j}}\eta^{(k)}_{ij}$. 
 $\eta_{ij}^{(k)}$ are complex coupling constants. Where $H_{k}^{0}=(h, H, A)$ are the three 
 physical neutral scalars. $H_{2}^{0}\equiv H^{0}$ plays the role of the Higgs boson in the SM. 
 We will neglect the fourth $t'$ quark effects by assuming small mixings 
 $|v_{t'}^{q}| \ll |v_{t}^{q}|$. It is then not difficult to find that 
 $Im(\tilde{C}_{q v}^{L}/v_{t}^{q}) = Im(\chi_{q}^{H^{0}}) C_{q v}^{H^{0}} $ ($v=\gamma, g$)  
  with $\chi_{q}^{H^{0}} = \sqrt{(m_{q}/4m_{b})} \eta_{qb'}^{(k)}\eta_{b'b}^{(k)}$, 
$C_{q\gamma}^{H^{0}}  =  \eta^{16/23}
[ -\frac{1}{3}E(y_{b'}) + \frac{8}{3} E(y_{b'}) (\eta^{-2/23} -1) ] $ and 
$C_{qg}^{H^{0}}  =  \eta^{14/23}E(y_{b'})$ 
(here $y_{b'} = m_{b'}^{2}/m_{H_{k}^{0}}^{2}$ 
and $\eta = \alpha_{s}(m_{W})/\alpha_{s}(m_{b})\simeq 0.56$). Of particular interest, 
the ratio $ C_{qg}^{H^{0}}/C_{q\gamma}^{H^{0}}$ is independent of the neutral Higgs masses. 
 To leading-order contributions, we have $C_{qg}^{H^{0}}/C_{q\gamma}^{H^{0}}\simeq -5.3$. 
 This leads to a parameter-independent maximal CP asymmetry 
 $a_{X_{q}\gamma}^{CP}|_{max} \simeq \mp 55\%$ and an enhanced branching ratio for 
 chromo-magnetic dipole decay $b\rightarrow sg$ at the level $B(b\rightarrow sg) \simeq 22\%$.

 When going back to the general case, we find that 
  for $B\rightarrow X_{s}\gamma$ processes when $m_{H^{0}}\simeq m_{b'}$ and $m_{H^{+}} 
  \simeq 200$GeV, $Im(\chi_{s}^{H^{+}}) \simeq 0$,  $Re(\chi_{s}^{H^{+}}) \simeq 4.5$, 
  $Im(\chi_{s}^{H^{0}}/v_{t}^{s}) \simeq \pm 7.0$ and 
$Re(\chi_{q}^{H^{0}}/v_{t}^{s}) \simeq -0.09$, 
  the resulting direct CP asymmetry is $A_{X_{s}\gamma}^{CP} \simeq \mp 27\%$. Here  
  $\mp54\%$ arise from the pure CP-violating scalar interactions and $\pm27\%$ from the 
interference of standard and 
  neutral scalar exchanges. Note that these two contributions have an opposite sign. 
  The resulting branching ratio of the chromo-magnetic dipole decay $b\rightarrow sg$ is 
  found to be $B(b\rightarrow sg)\simeq 7\%$. This value is very close to a preliminary 
  experimental bound $6.8\%$ (Ref. 14)  much 
larger than the SM predictions $B(b\rightarrow sg)\simeq 0.2\%$.
 This limit is increased to $9.0\%$ at $90\%$ C.L.
 if one uses more recent charmed baryon and chamonium yields presented in Refs. 15 and 16
and makes use of the relative $\Lambda_{c}$ and 
$\bar{\Lambda}_{c}$ yields given in Ref. 17 . Since there may exist large uncertainties when 
extracting the upper bound\cite{ALK}, we shall use the current bounds only as a 
reference. 
  
   Note that the following interesting case may occur, i.e., 
the direct CP asymmetry may become small but 
mixing-induced CP asymmetries suggested recently 
by Atwood, Gronau and Soni\cite{AGS} could be large.
The mixing-induced CP asymmetries require both the $B^{0}$ and $\bar{B}^{0}$  
    to be able to decay to a common final state. In the radiative B decays,  
    both should decay to states with the same photon
     helicity.  Let $A^{L,R}_{q}$ and $\bar{A}^{L,R}_{q}$ denote the decay amplitudes 
     of $B^{0}\rightarrow X^{0}_{q} \gamma_{L,R}$ and
      $\bar{B}^{0}\rightarrow X^{0}_{q} \gamma_{L,R}$, respectively.
We find that the amplitudes for emission of left- and right-handed photons in 
   $B^{0}\rightarrow X^{0}_{q} + \gamma$ are approximately  
   $A_{q}^{L} \propto v_{t}^{q}\bar{C}_{q\gamma}^{L} + 
v_{t}^{q}\chi_{q}^{H^{+}} C_{q\gamma}^{H^{+}} + 
   \chi_{q}^{H^{0}} C_{q\gamma}^{H^{0}}$ and 
 $  A_{q}^{R} \propto \tilde{\chi}_{q}^{H^{0}} C_{q\gamma}^{H^{0}}$ 
($\tilde{\chi}_{q}^{H^{0}}= \sqrt{(m_{q}/4m_{b})} \eta_{b'q}^{(k)}\eta_{bb'}^{(k)}$).
 It is seen that freedom of the parameters $\chi_{q}^{H^{+}}$, $\chi_{q}^{H^{0}}$ and 
 $\tilde{\chi}_{q}^{H^{0}}$ allows us to obtain solutions which satisfy 
 $|A_{q}^{L}| \simeq |A_{q}^{R}|$ with a maximal CP-violating phase.
It is easy to obtain a solution: 
$Im(\chi_{q}^{H^{0}}/v_{t}^{q}) \simeq Im\chi_{q}^{H^{+}}\simeq 0$ (i.e., new contributions to 
direct CP asymmetries will be negligible), $Re\chi_{q}^{H^{+}}\simeq 3.7$, $Re(\chi_{q}^{H^{0}}/v_{t}^{q}) 
 \simeq -2.2$ and $|\tilde{\chi}_{q}^{H^{0}}| \simeq 9.9$ for $m_{H^{+}} \simeq 200$GeV 
 and $m_{H^{0}}\simeq m_{b'}$.Such solutions will lead to maximal time-dependent mixing-induced 
CP asymmetry ${\cal A}_{q}(t)=(\Gamma(t) -\bar{\Gamma}(t)) / (\Gamma(t)+ \bar{\Gamma}(t))|_{max} 
\simeq  \pm \sin (\Delta m_{B} t)$ and time-integrated mixing-induced 
CP asymmetry ${\cal A}_{q}|_{max}\simeq  \pm 46\%$.
For this case, we find that the branching ratio for the chromo-magnetic dipole decay 
 $b\rightarrow sg$ is $B(b\rightarrow sg) \simeq 11\%$. Such a value was found\cite{HT} to be 
 interesting in understanding the recent observation \cite{CLEO3} of 
 large branching ratios concerning $\eta'$ yields in the charmless $B$ 
 decays. The values at this level were also known to be favorable to lower the theoretical
 predictions for the semileptonic branching ratio of $B\rightarrow X_{c} e \bar{\nu}$ decay 
 from $B(B\rightarrow X_{c} e \bar{\nu}) = 12\%$ to $B(B\rightarrow X_{c} e \bar{\nu}) = 10.8\%$.
 When taking $B(b\rightarrow sg) \simeq 6\%$, we find that ${\cal A}_{q}(t)|_{max} \simeq 
 \pm 0.87\sin (\Delta m_{B} t)$ and ${\cal A}_{q}|_{max} \simeq \pm 40\%$. 
  It is noted that the resulting ranges of the parameters  are consistent 
  with other constriants. In particular, it does not affect 
  very much the $B^{0}_{s}-\bar{B}^{0}_{s}$ mixing. Therefore, 
 mixing-induced CP asymmetries in radiative exclusive $B$ decays are sensitive to
 new physics that may only lead small contributions to the
 $B^{0}_{s}-\bar{B}^{0}_{s}$ mixing. It implies that to distinguish from new physics 
 phenomena that may have large contributions to the $B^{0}_{s}-\bar{B}^{0}_{s}$ mixing, 
 one may use, except by directly measuring this mixing, the mothed discussed 
 recently by London and Soni\cite{LS} by measuring the CP angle $\beta$ 
 via hadronic $b\rightarrow s$ penguin decays and comparing its value to that obtained 
 in $B\rightarrow \psi K_{S}$. 

 In concusion, though CP-averaged branching ratios of radiative B decays are roughly
  those expected from the SM, its radiative decay amplitudes
 can still receive larger contributions from new CP-violating interactions beyond 
  the SM. Such phenomena can be probed by the obervation of 
 sizable CP asymmetries in radiative B decays and an enhanced chromo-magnetic dipole 
decay $b\rightarrow sg$. The analyses presented here can easily be 
applied to many extensions of the SM, such as SU(2)$\times$ U(1) with vector-like quarks 
as well as nonminimal supersymmetric standard
 models\cite{KN,NSUSY}, where one should carefully analyze 
 all possible contributions together from various 
 interactions. It is of interest to note that direct CP asymmetries and mixing-induced 
 CP asymmetries in radiative $B$ decays as well as CP asymmetryies in hadronic 
 $b\rightarrow s$ penguin decays probe, to a large extent, different new physics effects
  and may compensate each other.

{\bf Acknowledgments}: The author would like to thank Lincoln Wolfenstein for 
reading the manuscript and for useful suggestions. This work was supported in 
part by the NSF of China under grant ($\#$ 19625514).

%\end{references}

\begin{thebibliography}{99}
%\begin{references}
%\bibitem[*]{byline}  
\bibitem{CLEO1} R. Ammar et. al. (CLEO Collaboration), Phys. Rev. 
Lett. {\bf 71}, 674 (1993).
\bibitem{CLEO2} M.S. Alam et al. (CLEO Collaboration), Phys. Rev. 
Lett. {\bf 74}, 2885 (1995)
\bibitem{TH} A. Ali and C. Greub, Phys. Lett. {\bf B259}, 182 (1991); 
Phys. Lett. B {\bf 361}, 146 (1995); 
K. Chetyrkin, M. Misiak and M. M\"{u}nz, Phys. Lett. B {\bf 400}, 206 (1997); 
A.J. Buras, A. Kwiatkowski and N. Pott, Phys. Lett. B {\bf 414}, 157 (1997); 
M. Ciuchini, G. Degrassi, P. Gambino and G.F. Giudice, hep-ph9710335; 
For most recent analyses and more references, see: A.L. Kagan and M. Neubert, 
CERN-TH/98-99, hep-ph/9805303.
\bibitem{CLEO3} B.H. Behrens et. al. (CLEO Collaboration), Phys. Rev. 
Lett. {\bf 80}, 3710 (1998).
\bibitem{WW1} L. Wolfenstein and Y.L. Wu, Phys. Rev. Lett. {\bf 73}, 2809 (1994).
\bibitem{WW2} Y.L. Wu and L. Wolfenstein, Phys. Rev. Lett. {\bf 73}, 1762 (1994).
\bibitem{KN} A.L. Kagan and M. Neubert, Phys. ev. {\bf D58}, 094012 (1998). 
\bibitem{JMS} J.M. Soares, Nucl. Phys. {\bf B367}, 575 (1991).
\bibitem{GHW} C. Greub, T. Hurth and D. Wyler, Phys. Lett. B {\bf 380}, 385 (1996); 
Phys. Rev. D {\bf 54}, 3350 (1996).
\bibitem{AI} H.M. Asatrian and A.N. Ioannissian, Phys. Rev. D {\bf 54}, 5642 (1996).
\bibitem{LO} B. Grinstein, R. Springer and M.B. Wise, Nucl. Phys. {\bf B339},
269 (1990); A.J. Buras, M. Misiak, M. M\"{u}nz and S. Pokorski, 
Nucl. Phys. B {\bf 424},  374 (1994).  
\bibitem{CS} T.P. Cheng and M. Sher, Phys. Rev. {\bf D35}, 3484 (1987).
\bibitem{HW} L.J. Hall and S. Weinberg, Phys. Rev. {\bf D48}, 979 (1993).
\bibitem{BSG} T.E., Coan et al, (CLEO Collaboration), Phys. Rev. Lett. {\bf 80}, 1150 (1998).
\bibitem{PD} P. Drell, CLNS-97-1521, hep-ex/9711020, in the Proceedings of the 18th International 
Symposium on Lepton-Photon Interactions, Hamburg, Germany, July 1997.
\bibitem{LG} L. Gibbons et al. (CLEO Collaboration), Phys. Rev. {\bf D56}, 3783 (1997).
\bibitem{DC} D. Cinabro et al. (CLEO Collaboration), Conference contribution CLEO-CONF 94-8, 
submitted to the International Conference on High Energy Physics, Glasgow, Scotland, July 1994.
\bibitem{ALK} A.L. Kagan, UCTP 107.98, hep-ph/9806266, to appear in the Proc. of the 7th Intern. 
Symposium on Heavy Flavor Physics, Santa Barbara, California, July 7-11, 1997. 
\bibitem{AGS} D. Atwood, M. Gronau and A. Soni,  Phys. Rev. Lett. {\bf 79}, 185 (1997).
\bibitem{HT} W.-S. Hou and B. Tseng,  Phys. Rev. Lett. {\bf 80}, 434 (1998); 
A.L. Kagan and A.A. Petrov, UCHEP-97/27, hep-ph/9707354.
\bibitem{LS} D. London and A. Soni, Phys.Lett. B {\bf 407}, 61 (1997).
\bibitem{NSUSY} F. Gabbiani, E. Gabrielli, A. Masiero and L. Silvestrini, 
Nucl. Phys. B {\bf 477}, 321 (1996).
\end{thebibliography}
\end{document}